\documentstyle[12pt]{article}
\makeatletter
\def\emline#1#2#3#4#5#6{%
       \put(#1,#2){\special{em:moveto}}%
       \put(#4,#5){\special{em:lineto}}}
\def\newpic#1{}

\def\hybrid{\topmargin 0pt      \oddsidemargin 0pt
        \headheight 0pt \headsep 0pt
        \textwidth 6.25in       
        \textheight 9.5in       
        \marginparwidth 0.0in
        \parskip 5pt plus 1pt   \jot = 1.5ex}
\catcode`\@=11
\def\marginnote#1{}

\newcount\hour
\newcount\minute
\newtoks\amorpm
\hour=\time\divide\hour by60
\minute=\time{\multiply\hour by60 \global\advance\minute by-\hour}
\edef\standardtime{{\ifnum\hour<12 \global\amorpm={am}%
        \else\global\amorpm={pm}\advance\hour by-12 \fi
        \ifnum\hour=0 \hour=12 \fi
        \number\hour:\ifnum\minute<10 0\fi\number\minute\the\amorpm}}
\edef\militarytime{\number\hour:\ifnum\minute<10 0\fi\number\minute}

\def\draftlabel#1{{\@bsphack\if@filesw {\let\thepage\relax
   \xdef\@gtempa{\write\@auxout{\string
      \newlabel{#1}{{\@currentlabel}{\thepage}}}}}\@gtempa
   \if@nobreak \ifvmode\nobreak\fi\fi\fi\@esphack}
        \gdef\@eqnlabel{#1}}
\def\@eqnlabel{}
\def\@vacuum{}
\def\draftmarginnote#1{\marginpar{\raggedright\scriptsize\tt#1}}

\def\draftlabel#1{{\@bsphack\if@filesw {\let\thepage\relax
   \xdef\@gtempa{\write\@auxout{\string
      \newlabel{#1}{{\@currentlabel}{\thepage}}}}}\@gtempa
   \if@nobreak \ifvmode\nobreak\fi\fi\fi\@esphack}
        \gdef\@eqnlabel{#1}}
\def\@eqnlabel{}
\def\@vacuum{}
\def\draftmarginnote#1{\marginpar{\raggedright\scriptsize\tt#1}}

\def\draft{\oddsidemargin -.5truein
        \def\@oddfoot{\sl preliminary draft \hfil
        \rm\thepage\hfil\sl\today\quad\militarytime}
        \let\@evenfoot\@oddfoot \overfullrule 3pt
        \let\label=\draftlabel
        \let\marginnote=\draftmarginnote
   \def\@eqnnum{(\theequation)\rlap{\kern\marginparsep\tt\@eqnlabel}%
\global\let\@eqnlabel\@vacuum}  }

\def\numberbysection{\@addtoreset{equation}{section}
        \def\theequation{\thesection.\arabic{equation}}}

\def\underline#1{\relax\ifmmode\@@underline#1\else
        $\@@underline{\hbox{#1}}$\relax\fi}

\def\titlepage{\@restonecolfalse\if@twocolumn\@restonecoltrue\onecolumn
     \else \newpage \fi \thispagestyle{empty}\c@page\z@
        \def\thefootnote{\fnsymbol{footnote}} }

\def\endtitlepage{\if@restonecol\twocolumn \else  \fi
        \def\thefootnote{\arabic{footnote}}
        \setcounter{footnote}{0}}  
\relax

\numberbysection
\hybrid

\def\beq{\begin{equation}}
\def\eeq{\end{equation}}

\def\t{\tau}
\def\l{\lambda}

\newdimen\Squaresize \Squaresize=30pt
\newdimen\Thickness \Thickness=0.5pt

\def\Square#1{\hbox{\vrule width \Thickness
   \vbox to \Squaresize{\hrule height \Thickness\vss
      \hbox to \Squaresize{\hss#1\hss}
   \vss\hrule height\Thickness}
\unskip\vrule width \Thickness}    
\kern-\Thickness}                  

\def\Vsquare#1{\vbox{\Square{$#1$}}\kern-\Thickness}

\begin{document}

\begin{titlepage}

\title{Tau-function for discrete
sine-Gordon equation and quantum $R$-matrix}

\author{A.Zabrodin\thanks{Joint Institute of
Chemical Physics, Kosygina str. 4, 117334,
Moscow, Russia and ITEP, 117259, Moscow, Russia}}
\date{September 1998}
\maketitle

\begin{abstract}

We prove that the $\tau$-function of the
integrable discrete sine-Gordon model apart from the
"standard" bilinar identities obeys a number of "non-standard"
ones. They can be combined into a bivector
3-dimensional difference equation which is shown to
contain Hirota's difference analogue of the
sine-Gordon equation and both auxiliary linear problems for it.
We observe that this equation is
most naturally written in terms
of the quantum $R$-matrix for the $XXZ$ spin chain and looks then
like a relation of the "vertex-face correspondence" type.

\end{abstract}


\end{titlepage}

\section{Introduction}

Integrable analogues of soliton equations in discrete
space-time attract much attention for they accumulate
developments of the theory of integrable models in
the most pure form
and therefore serve as a good source of new ideas.
The activity in the field of
partial difference integrable equations was initiated
by Hirota \cite{HirotaKdV}-\cite{Hirota} who proposed
discrete analogues of the KdV,
the Toda chain, the sine-Gordon (SG) equations, etc
(actually of almost all interesting soliton equations).
The systematic studies were carried on in the
works \cite{Miwa,DJM},\,\cite{CapelNij},\,\cite{SS}.
Among other things, they allowed us to understand
why Hirota's "empirical" method did really
work and generated true discrete
analogues of soliton equations.

An important contribution to this understanding
is Miwa's interpretation \cite{Miwa} of
the discrete time flows, in which discretized
integrable equations are treated as members
of the same infinite hierarchy as their continuous counterparts.
The general method
to produce discrete soliton equations from continuous ones \cite{DJM}
relies on the Miwa transformation -- a special change of
(in general infinitely many) independent
variables connecting continuous and discrete time flows.
A fundamental new ingredient making all this really meaningful
is the {\it $\tau$-function} of an infinite
hierarchy of integrable equations \cite{tau}.
Briefly speaking,
the $\tau$-function solves all equations of
the hierarchy simultaneously.

In few words, the main message of the present paper
sounds as follows:
without any quantization
the $\tau$-function is already
on friendly terms with the quantum $R$-matrix,
a famous object specific for
algebraic structure of quantum integrable models.
What we actually have in mind
can be most easily demonstrated for
the discrete SG model on which we concentrate in the sequel.

In its original form, Hirota's discrete analogue of the SG
equation \cite{HirotaSG} reads
\begin{eqnarray}
&&\sin \big (\phi (u,v)+\phi (u+1, v+1)-
\phi (u+1, v)- \phi (u,v+1) \big )
\nonumber \\
&=&\kappa \sin \big (\phi (u,v)+\phi (u+1, v+1)+
\phi (u+1,v)+ \phi (u,v+1)\big )\,,
\label{H1}
\end{eqnarray}
where $u,v$ are lattice light cone variables and $\kappa$
is a constant. Recently this equation was shown \cite{BP} to
play a remarkable role in the discrete analogue
of continuous two-dimensional differential geometry.
The Faddeev-Volkov version \cite{FadVol} of this
equation,
\begin{eqnarray}
&&\psi (u,v+1)\psi (u+1, v+1)-
\psi (u,v)\psi (u+1, v)
\nonumber \\
&=&\kappa \psi (u+1,v)\psi (u+1, v+1)-
\kappa \psi (u,v)\psi (u, v+1),
\label{FV}
\end{eqnarray}
is sometimes more convenient to deal with. It
can be obtained from eq.\,(\ref{H1}) via the substitution
$\psi (u,v)=\mbox{exp} \bigl ( 2i e^{\pi i v} \phi (u,v)\bigr )$.
It was shown \cite{FadVol}, \cite{Volkov}
that different continuum limits of eq.\,(\ref{FV})
yield SG and KdV equations.
Moreover, eq.\,(\ref{FV} itself is equivalent to Hirota's discrete
analogue of the KdV equation (see eq.\,(\ref{DKdV}) below).
It should be also noted that classical
and quantum integrable versions of the SG equation on
the space lattice with continuous time were suggested in
\cite{IK}.

The structure of discretized soliton
equations becomes much more transparent if to pass to
the $\tau$-function. From more practical point of view,
the reformulation in terms of the $\tau$-function makes
somewhat misteriously looking
substitutions easy to guess.

To be more specific, let us
introduce the $\tau$-function for eq.\,(\ref{FV}).
The idea of Hirota's approach is to represent the
equation as a
consequence of a 3-term bilinear identity
for the $\tau$-function
$\tau _{n}(u,v)$ \cite{Hirota}, which is simpler and
in a sense more fundamental:
\beq
\label{3term}
\tau _{n}(u\!+\!1,v)\tau _{n+1}(u,v\!+\!1)
-\kappa \tau _{n}(u,v\!+\!1)\tau _{n+1}(u\!+\!1,v)
=(1\!-\!\kappa )\tau _{n}(u\!+\!1,v\!+\!1)\tau _{n+1}(u,v).
\eeq
Here $n$ is a third discrete variable.
This equation is valid in the general 2D Toda lattice
hierarchy \cite{UT} discretized along the lines of \cite{DJM}.
The subhierarchy with the additional
constraint
\beq
\label{red}
\tau _{n+2}(u,v)=
\tau _{n}(u,v)
\eeq
is the (discrete) SG hierarchy or, equivalently, 2-periodic
2D Toda lattice \cite{UT}. Taking into account this constraint,
it is convenient to write
\beq
\tau _{0}(u,v)\equiv \tau (u,v),
\;\;\;\;\;\;\;\;
\tau _{1}(u,v)\equiv \hat \tau (u,v)\,.
\label{red1}
\eeq
Setting
$$
\psi (u,v)=\frac{\hat \tau (u,v)}{\tau (u,v)}\,,
$$
we immediately see that eq.\,(\ref{FV}) follows from
eq.\,(\ref{3term}). Higher equations of the hierarchy include
more discrete variables.

In this paper we prove a number of additional identities
obeyed by the $\tau$-function of the SG hierarchy.
Among them there are discrete KdV and Toda chain equations in
bilinear form. Combining them
into one matrix identity, we recover that it can be most
naturally written in terms of the quantum $R$-matrix of the
$XXZ$ spin chain. The identity then looks similarly to
what is called "the vertex-face correspondence" between
models of statistical mechanics on the lattice
(see \cite{Baxter}--\cite{vertex-face}).
"Vertex" and "face" types of
Boltzmann weights are related through "intertwining vectors".
In our case components of the
"intertwining vectors" are identified with the $\tau$-function
evaluated at different space-time points.

Treated as an equation, not just an identity, the
above mentioned relation contains two linear problems
and, therefore, implies their compatibility.
In its turn, this implies a zero
curvature condition on the lattice for certain matrix functions
$L$, $M$ (the $L$-$M$ pair), which is equivalent to the
discretized SG equation.
In this way we reproduce the results of our
previous papers \cite{Z}, where the $L$-$M$ pair has been
written in terms of the quantum $R$-matrix.

\section{General form of 3-term bilinear equations}

Here we briefly review some elements of the bilinear
formalism for difference soliton equations.
For more details see e.g. \cite{DJM},\,\cite{Zab}.

The key object of
the bilinear formalism is {\it $\t$-function}.
We start with the $\tau$-function of the 2D Toda lattice
hierarchy \cite{UT}.
The $\t$-function depends on two infinite sets
of continuous times ${\bf t}=\{t_1, t_2, t_3, \ldots \}$,
$\bar {\bf t}=\{\bar t_1, \bar t_2, \bar t_3, \ldots \}$
(which we call "positive" and "negative" respectively) and
on a discrete "time" $n\in {\bf Z}$. For our purpose we do not
need the "negative" times, so let us write the $\t$-function
as $\tau _{n}(t_1 , t_2 , t_3 , \ldots )
\equiv \t _{n}({\bf t})$, where all "negative" times (not shown
explicitly) are supposed to be fixed.

First of all we recall how to introduce discrete time flows
via Miwa's transformation.
We use the standard short hand notation
$$
\tau ({\bf t}\pm k[z])= \tau (t_1 \pm kz, t_2 \pm \frac{1}{2}kz^2,
t_3 \pm \frac{1}{3}kz^3, \ldots )\,,
\;\;\;\;\; k\in {\bf Z}\,.
$$
The discrete time $k$ is introduced by the definition
\beq
\label{p}
\tau _{n}^{k} =\tau _n({\bf t}^{(0)} -k[\lambda _{p}^{-1}])\,,
\eeq
where ${\bf t}^{(0)}$ stands for constant "background" values
of the continuous times and $\lambda _{k}\in {\bf C}$
is a parameter of the flow
sometimes called {\it Miwa's variable}.
The $\l _k$ should be thought of as a continuous "label"
associated with the discrete flow.
From physical point of view $\l _{k}^{-1}$ serves as a kind of
lattice spacing in the $k$-direction, as it is seen
from (\ref{p}). One may introduce arbitrary number
of independent discrete flows $a,b,c, \ldots $
according to the same prescription:
\beq
\label{abc}
\tau _{n}^{abc\ldots } =
\tau _n({\bf t}^{(0)} -a[\lambda _{a}^{-1}]
-b[\lambda _{b}^{-1}] -c[\lambda _{c}^{-1}] -\ldots )\,.
\eeq
It is clear from (\ref{abc}) that if $\lambda _a =\lambda _b$,
then the flows $a$ and $b$ must be identified. This obvious
remark will be important in what follows.

From now on the discrete times will be written
as arguments of the $\t$-function
(like $\t _n(a,b,c, \ldots)$) rather than upper indices.
Since the continuous times $\{t_i\}$
will not enter the play any more,
this should not lead to a confusion.
For any two discrete flows $a,b$ we put
$$
\l _{ab}\equiv \l _a -\l _b \,,
\;\;\;\;\;\;
\l _{ab}=-\l _{ba}\,.
$$

As a function of the discrete times, the $\t$-function
obeys a number
of bilinear partial difference equations.
Let us recall their general form.
For any triplet $\{ a,b,c\}$ of discrete flows one
has the following 3-term bilinear
equation \cite{Hirota},\,\cite{Miwa}
for the $\t$:

\begin{eqnarray}
&&\l _{bc}\t (a+1, b,c)\t (a, b+1, c+1)
\nonumber \\
&+&\l _{ca}\t (a, b+1,c)\t (a+1, b, c+1)
\nonumber \\
&+&\l _{ab}\t (a, b,c+1)\t (a+1, b+1, c)=0\,.
\label{tripl}
\end{eqnarray}

\noindent
For any quadruplet $\{ a,b,c,d \}$
one has another 3-term bilinear
equation:

\begin{eqnarray}
&&\l _{ad}\l _{bc}\t (a+1, b,c, d+1)\t (a, b+1, c+1,d)
\nonumber \\
&+&\l _{bd}\l _{ca}\t (a, b+1,c, d+1)\t (a+1, b, c+1,d)
\nonumber \\
&+&\l _{cd}\l _{ab}\t (a, b,c+1, d+1)\t (a+1, b+1, c,d)=0\,.
\label{quadrupl}
\end{eqnarray}

\noindent
There are links between
eqs.\,(\ref{tripl}), (\ref{quadrupl})
in both directions.
On the one hand, eq.\,(\ref{tripl}) is a particular case
of (\ref{quadrupl}) when $\l _{d} \to \infty$. (According
to (\ref{p}), this limit means that the dependence
on $d$ in the $\t$ disappears.) On the other hand,
eq.\,(\ref{quadrupl}), though linearly independent
of eqs.\,(\ref{tripl}), is an {\it algebraic consequense} of
equations of the type (\ref{tripl}) written for
the triplets $\{a,b,c\}$, $\{a,b,d\}$ and $\{a,c,d\}$.
In this sense all what we are going to derive in the
sequel is implicitly contained in equations of the
type (\ref{tripl}).

Although the discrete time $n$ was distinguished from
the very beginning, we can treat it on equal footing
with the other ones, assigning to it
the "label" $\lambda _n =0$.
In particular, one can write equations
similar to (\ref{tripl}), (\ref{quadrupl}) for the triplet
$(n,a,b)$ and the quadruplet $(n,a,b,c)$
with $\lambda _n =0$.

\paragraph{Example.}
For illustrative purpose we recall the explicit form
of the $N$-soliton $\t$-function.
Fix complex parameters $p_1 , p_2 , \ldots p_N$, $q_1 , q_2 ,
\ldots , q_N$ and $\alpha _1 , \alpha _2 , \ldots , \alpha _N$.
Then the $\t$-function
(as a function of discrete variables $a_1 , a_2 , a_3 , \ldots$)
is given by the determinant formula \cite{OHTI}
$$
\t _n (a_1, a_2 , \ldots )=
\det \left [ \delta _{ij} +
\frac{\alpha _j}{p_j -q_i }\left ( \frac{p_j}{q_j}\right )^n
\prod _{l}\left ( \frac{p_j -\lambda _{a_l}}{q_j -\lambda _{a_l}}
\right )^{a_l}\right ]_{1\leq i,j \leq N}.
$$
It obeys all the bilinear relations listed above.

\section{Bilinear equations for the $\tau$-function
of the discrete SG model}

Let us specify the general formulas for the case
of the discrete SG model.
Fix two discrete flows $u,v$ with the corresponding
parameters $\lambda _u$, $\lambda _v$.
For brevity we set $\mu \equiv \l _{u}$,
$\nu \equiv \l _{v}$.
Let $\tau _n(u,v)$ be the $\tau$-function.
As it was already said in the Introduction,
the SG hierarchy is defined by the
{\it reduction condition} (\ref{red}).
(In the example given at the end of the previous section
this reduction is equivalent to the constraints
$p_i +q_i =0$, $i=1,\ldots , N$.)

Using the notation (\ref{red1}), we rewrite
eq.\,(\ref{tripl}) for the triplet $(n,u,v)$ as
the pair of coupled equations for
$\t$ and $\hat \t$:
\beq
\nu \tau (u+1,v)\hat \tau (u,v+1)
-\mu \hat \tau (u+1,v)\tau (u,v+1)=
(\nu -\mu )\hat \tau (u,v)\tau (u+1,v+1)\,,
\label{03a}
\eeq
\beq
\nu \hat \tau (u+1,v) \tau (u,v+1)
-\mu \tau (u+1,v)\hat \tau (u,v+1)=
(\nu -\mu )\tau (u,v)\hat \tau (u+1,v+1)\,,
\label{03b}
\eeq
They can be rearranged to the following equivalent form:
\beq
\mu \hat \tau (u,v) \tau (u+1,v+1)
+\nu \tau (u,v)\hat \tau (u+1,v+1)=
(\nu +\mu )\tau (u,v+1)\hat \tau (u+1,v)\,,
\label{04a}
\eeq
\beq
\mu \tau (u,v)\hat \tau (u+1,v+1)
+\nu \hat \tau (u,v)\tau (u+1,v+1)=
(\nu +\mu )\hat \tau (u,v+1)\tau (u+1,v)\,.
\label{04b}
\eeq
The Faddeev-Volkov equation (\ref{FV}) for
$\psi (u,v)=\hat \t (u,v)/\t (u,v)$
holds with $\kappa =\mu /\nu$.
As is already clear from the previous section, the
parameters $\mu$ and $\nu$ play the role of inverse lattice spacings
in the chiral directions $u,v$ respectively.

There is a relation connecting values of $\t$ and $\hat \t$
along the $u$-direction (or the $v$-direction).
To derive it, introduce a
third discrete variable $u'$ with Miwa's variable $\mu '$. Then equations of
the form (\ref{04a}), (\ref{04b}) are valid for $u'$ in place of $v$ (and
$\mu '$ in place of $\mu$) while $v$ is not shifted and enters as a
parameter.  Recalling the remark after eq.\,(\ref{abc}), we should identify
$u'$ with $u$ as $\mu '=\mu$; this yields the relation
\beq
\tau (u-1, v)\hat \tau (u+1, v)
+\hat \tau (u-1, v)\tau (u+1, v)
=2\tau (u, v)\hat \tau (u, v)\,.
\label{05}
\eeq
We have obtained basic bilinear equations of the discrete
SG model.

Besides the basic equations given above, the $\t$-function
obeys some other 3-term bilinear identities which follow from the
basic ones.
Here we mention three of them:

\begin{eqnarray}
&& (\mu +\nu )\tau (u-1,v+1)\tau (u+1,v)
+(\mu -\nu )\tau (u+1,v+1)\tau (u-1,v)
\nonumber \\
&=&2\mu \tau (u,v)\tau (u,v+1)\,,
\label{06a}
\end{eqnarray}

\begin{eqnarray}
&& (\mu +\nu )\tau (u-1,v+1)\hat \tau (u+1,v)
-(\mu -\nu )\tau (u+1,v+1)\hat \tau (u-1,v)
\nonumber \\
&=&2\nu \tau (u,v)\hat \tau (u,v+1)\,,
\label{06b}
\end{eqnarray}

\begin{eqnarray}
&& (\mu \!+\!\nu )^2\tau (u\!-\!1,v\!+\!1) \tau (u\!+\!1,v\!-\!1)
-(\mu \!-\!\nu )^2\tau (u\!+\!1,v\!+\!1)\tau (u\!-\!1,v\!-\!1)
\nonumber \\
&=&4\mu \nu (\tau (u,v))^2.
\label{06c}
\end{eqnarray}

\noindent
Their derivation
from the basic equations is given in the Appendix. Note also
that since $\t$ and $\hat \t$ differ only by the shift of
the hidden variable $n$ and $\t _{n+2}=\t _{n}$, the exchange
$\t \leftrightarrow \hat \t$ in
eqs.\,(\ref{06a})--(\ref{06c})
automatically yields true identities which we do not
write down explicitly.

We note in passing that eqs.\,(\ref{06a}) and (\ref{06c})
provide bilinearization of Hirota's discrete KdV equation
and 1D Toda chain in discrete time respectively. Indeed,
it follows from (\ref{06a}) that
$$
U(u,v)=\frac{\tau (u,v)\tau (u+1, v+1)}
{\tau (u+1,v)\tau (u, v+1)}
$$
solves the discrete KdV equation \cite{HirotaKdV}
\beq
U(u+1,v+1)-U(u, v)=
\frac{1+\kappa}{1-\kappa}
\Bigl (U^{-1}(u, v+1)-U^{-1}(u+1, v)\Bigr )
\label{DKdV}
\eeq
with $\kappa =\mu /\nu$.
Note also that substituting
$$
U(u,v)=\frac{\psi (u,v+1)-\kappa \psi (u+1,v)}{(1-\kappa )
\psi (u,v)}\,,
$$
we get the Faddeev-Volkov equation (\ref{FV}) for $\psi$.

Further, eq.\,(\ref{06c}) implies that
$$
\varphi (u,v)=\mbox{log}
\frac{\t (u+1, v+1)}{\t (u,v)}
$$
obeys the equation
\begin{eqnarray}
\label{1DTC}
&& \exp \Bigl ( \varphi (u+1, v-1)+
\varphi (u-1, v+1)
-2\varphi (u, v)\Bigr )
\nonumber \\
&=& \frac{
1+g \exp \bigl (\varphi (u+1, v+1)-\varphi (u, v)\bigr )}
{1+g \exp \bigl (\varphi (u, v)-\varphi (u-1, v-1)\bigr )}
\end{eqnarray}
with $g=4\mu \nu /(\mu -\nu )^2$, which after passing to the
"laboratory" coordinates $\frac{1}{2}(u\pm v)$ coincides with
the 1D Toda chain equation in discrete time
\cite{HirotaTC} (see also \cite{Suris}). We stress that all this
holds for {\it the same} $\t$-function $\t (u,v)$ of the discrete
SG equation.

\section{3D difference equations for the SG
$\tau$-function}

Let us turn to bilinear equations for the SG $\tau$-function
which include more discrete variables. In some sense they are
discrete counterparts of higher equations of the SG hierarchy.
Here we consider the case of one such "extra variable" $w$
introduced according to (\ref{abc}).
Set $\l _{w}=\zeta$. The $\tau$-function
is now written as $\t _0 (u,v,w)=\tau (u,v,w)$
and $\t _1 (u,v,w)=\hat \tau (u,v,w)$.
It satisfies the equations

\beq
\label{e1}
\begin{array}{ll}
&(\zeta -\mu )\hat \tau (u,v,w)\tau (u+1,v, w+1)\\
=&\zeta \tau (u+1,v,w)\hat \tau (u,v, w+1)\\
-&\mu \hat \tau (u+1,v,w)\tau (u,v, w+1)\,,
\end{array}
\eeq

\beq
\label{e2}
\begin{array}{ll}
&(\zeta -\nu )\hat \tau (u,v,w)\tau (u,v+1, w+1)\\
=&\zeta \tau (u,v+1,w)\hat \tau (u,v, w+1) \\
-&\nu \hat \tau (u,v+1,w)\tau (u,v, w+1)
\end{array}
\eeq

\noindent
(and the same with the change $\tau \leftrightarrow \hat \tau$).
They are nothing else than eq.\,(\ref{03a}) written for the
triplets $(p,u,w)$, $(p,v,w)$ respectively.
Similarly, the analogues of eqs.\,(\ref{04a}), (\ref{04b}) are

\beq
\label{e3}
\begin{array}{ll}
&(\zeta +\mu )\tau (u+1,v,w)\hat \tau (u,v,w+1)\\
=&\mu \hat \tau (u+1,v,w+1) \tau (u,v,w)\\
+&\zeta \tau (u+1,v, w+1)\hat \tau (u,v,w)\,,
\end{array}
\eeq

\beq
\label{e4}
\begin{array}{ll}
&(\zeta +\nu )\tau (u,v+1,w)\hat \tau (u,v,w+1)\\
=&\nu \hat \tau (u,v+1,w+1) \tau (u,v,w)\\
+&\zeta \tau (u,v+1, w+1)\hat \tau (u,v,w)
\end{array}
\eeq

\noindent
(and again the same with the change
$\tau \leftrightarrow \hat \tau$).

Eq.\,(\ref{tripl}) for the triplet $(u,v,w)$ gives
\beq
\begin{array}{ll}
&(\zeta -\mu ) \tau (u,v+1,w)\tau (u+1,v, w+1) \\
-&(\zeta -\nu)\tau (u+1,v,w) \tau (u,v+1, w+1)\\
+&(\mu -\nu ) \tau (u,v,w+1)\tau (u+1,v+1, w)=0\,.
\end{array}
\label{e5}
\eeq
Eq.\,(\ref{quadrupl}) for the quadruplet $(p,u,v,w)$ gives
\beq
\begin{array}{ll}
&\nu(\zeta -\mu ) \hat \tau (u,v+1,w)\tau (u+1,v, w+1) \\
-&\mu (\zeta -\nu)\hat \tau (u+1,v,w) \tau (u,v+1, w+1)\\
+&\zeta (\mu -\nu ) \hat \tau (u,v,w+1)\tau (u+1,v+1, w)=0\,.
\end{array}
\label{e6}
\eeq

In the next section we need two more complicated bilinear
equations:

\beq
\label{e7}
\begin{array}{ll}
&(\mu +\nu )(\zeta -\mu ) \tau (u-1,v+1,w)\tau (u+1,v, w+1) \\
+&(\mu -\nu)(\zeta +\mu)\tau (u-1,v,w+1) \tau (u+1,v+1, w)\\
=&2\mu (\zeta -\nu )  \tau (u,v,w)\tau (u,v+1, w+1)\,,
\end{array}
\eeq

\vspace{0.4cm}

\beq
\label{e8}
\begin{array}{ll}
&(\mu +\nu )(\zeta -\mu )
\hat \tau (u-1,v+1,w)\tau (u+1,v, w+1) \\  \\
-&(\mu -\nu)(\zeta +\mu)\tau (u-1,v,w+1) \hat \tau (u+1,v+1, w)\\ \\
=&\displaystyle{\frac{
2\nu (\zeta ^2-\mu ^2)}{\zeta +\nu }} \,
\hat \tau (u,v,w)\tau (u,v+1, w+1) \\  \\
+&\displaystyle{\frac{
2\zeta (\nu ^2-\mu ^2)}{\zeta +\nu }} \,
\tau (u,v,w)\hat \tau (u,v+1, w+1)
\end{array}
\eeq

\noindent
(and the same with $\t \leftrightarrow \hat \t$).
They can be derived from eqs.\,(\ref{e1})--(\ref{e6})
in the manner similar to how eqs.\,(\ref{06a})--(\ref{06c})
have been derived from the basic equations (\ref{03a})--(\ref{04b}).
The derivation is outlined in the Appendix.

\section{$R$-matrix form of the 3D bilinear identities}

Our goal in this section is to combine the identities
(\ref{e7}), (\ref{e8}) into one compact equation suggesting
parallels with lattice
integrable models of statistical mechanics.

Let us introduce the matrix of Boltzmann weights
of the 6-vertex model:
\beq
\label{R}
\mbox{{\sf R}}=\left (
\begin{array}{cccc}
\mbox{{\sf a}}&0&0&0\\
0&\mbox{{\sf b}}&\mbox{{\sf c}}&0\\
0&\mbox{{\sf c}}&\mbox{{\sf b}}&0\\
0&0&0&\mbox{{\sf a}}
\end{array}
\right )
\eeq
and call it $R$-matrix. At the moment
we do not imply any specific parametrization of the
matrix elements. Assuming the standard block structure
of the $R$-matrix, we say that it acts in the tensor
product $V_1 \otimes V_2$ of two 2-dimensional spaces
isomorphic to ${\bf C}^2$:
$\mbox{{\sf R}}\in \mbox{End} (V_1 \otimes V_2) $,
$V_1 =V_2 ={\bf C}^2$. In other words,
$$
\mbox{{\sf R}}=
\frac{1}{2}(\mbox{{\sf a}}+\mbox{{\sf b}})I\otimes I
+\frac{1}{2}(\mbox{{\sf a}}-\mbox{{\sf b}})
\sigma _3\otimes \sigma _3
+\frac{1}{2}\mbox{{\sf c}}(\sigma _1\otimes \sigma _1
+\sigma _2\otimes \sigma _2)\,.
$$
(Here $I$
is the unit matrix and $\sigma _i$ are Pauli matrices.)
Hence the $R$-matrix acts
to tensor product of the vectors
$X= \left (\begin{array}{c}x_1\\x_2 \end{array}\right )\in V_1$,
$Y= \left (\begin{array}{c}y_1\\y_2 \end{array}\right )\in V_2$
as follows:
$$
\mbox{{\sf R}}\,X\otimes Y=
\mbox{{\sf R}}\left (\begin{array}{c}
x_1y_1\\x_2y_1\\x_1y_2\\x_2y_2\end{array}\right )=
\left (\begin{array}{c}
\mbox{{\sf a}}x_1y_1\\\mbox{{\sf b}}x_2y_1+
\mbox{{\sf c}}x_1y_2
\\\mbox{{\sf c}}x_2y_1+\mbox{{\sf b}}x_1y_2\\
\mbox{{\sf a}}x_2y_2\end{array}\right ).
$$

It is easy to verify that eqs.\,(\ref{e7}), (\ref{e8})
are equivalent to the relation
\beq
\label{rhir}
\begin{array}{ll}
&\mbox{\large {\sf R}}
\left (\begin{array}{c}\tau (u,v,w)
\\ \\ \hat \tau (u,v,w)
\end{array}\right )
\otimes
\left (\begin{array}{c}\tau (u,v+1,w+1)
\\ \\ -\hat \tau (u,v+1,w+1)
\end{array}\right )  \\ &\\
=&
\alpha
\left (\begin{array}{c}\tau (u-1,v+1,w)
\\ \\ \hat \tau (u-1,v+1,w)
\end{array}\right )
\otimes
\left (\begin{array}{c}\tau (u+1,v,w+1)
\\ \\ -\hat \tau (u+1,v,w+1)
\end{array}\right )  \\ &\\
+&
\beta
\left (\begin{array}{c}\tau (u+1,v+1,w)
\\ \\ -\hat \tau (u+1,v+1,w)
\end{array}\right )
\otimes
\left (\begin{array}{c}\tau (u-1,v,w+1)
\\ \\ \hat \tau (u-1,v,w+1)
\end{array}\right ),
\end{array}
\eeq

\noindent
where the $R$-matrix has elements
\beq
\begin{array}{l}
\mbox{{\sf a}}=2\mu (\zeta ^2 -\nu ^2)\\ \\
\mbox{{\sf b}}=2\nu (\zeta ^2 -\mu ^2)\\ \\
\mbox{{\sf c}}=2\zeta (\mu ^2 -\nu ^2)
\end{array}
\eeq
and $\alpha $, $\beta $ are scalar coefficients:

\beq
\begin{array}{l}
\alpha =
\frac{1}{2} (\mbox{{\sf a}}+\mbox{{\sf b}}+\mbox{{\sf c}})
=(\mu +\nu )(\nu +\zeta )(\zeta -\mu )\\ \\
\beta =
\frac{1}{2}(\mbox{{\sf a}}-\mbox{{\sf b}}-\mbox{{\sf c}})
=(\mu -\nu )(\nu +\zeta )(\zeta +\mu )\,.
\end{array}
\eeq

\noindent
After a proper rescaling and extracting a common
multiplier, {\sf R} becomes the standard trigonometric $R$-matrix
with spectral parameter $\zeta$:
$\mbox{{\sf R}}=2\mu \nu \zeta
\mbox{{\sf R}}_{\kappa}(\zeta /\mu )$,
where $\kappa =\mu /\nu$ and
\beq
\label{Rtrig}
\mbox{{\sf R}}_{\kappa}(z)=
\left (
\begin{array}{ccccccc}
\kappa z \!-\! \kappa ^{-1}z^{-1} && 0 && 0 && 0\\
0 && z\!-\!z^{-1} && \kappa \!-\!\kappa ^{-1} && 0  \\
0 && \kappa \!-\!\kappa ^{-1} && z\!-\!z^{-1} && 0  \\
0 && 0 && 0 && \kappa z \!-\! \kappa ^{-1}z^{-1}
\end{array}
\right )
\eeq
is the $R$-matrix of the $XXZ$-type with the
quantum deformation parameter $\kappa$
and the spectral parameter $z$. It obeys the Yang-Baxter
equation
\beq
\label{YB}
\mbox{{\sf R}}_{\kappa}^{12}(z_1/z_2)
\mbox{{\sf R}}_{\kappa}^{13}(z_1/z_3)
\mbox{{\sf R}}_{\kappa}^{23}(z_2/z_3)=
\mbox{{\sf R}}_{\kappa}^{23}(z_2/z_3)
\mbox{{\sf R}}_{\kappa}^{13}(z_1/z_3)
\mbox{{\sf R}}_{\kappa}^{12}(z_1/z_2)\,,
\eeq
where
$\mbox{{\sf R}}_{\kappa}^{ij}$ acts as
$\mbox{{\sf R}}_{\kappa}$ in the tensor product of
$i$-th and $j$-th spaces and as the unit matrix
in the remaining space.

To clarify the meaning of eq.\,(\ref{rhir}), we rewrite it as follows:
\begin{eqnarray}
\label{rhir1}
&&\mbox{{\sf R}}_{\kappa}(z) \Phi (u,v)\otimes \Psi (u,v+1)
\nonumber \\
&=& \!\!\alpha (z)\Phi (u\!-\!1,v\!+\!1)\otimes
\Psi (u\!+\!1,v)
+\beta (z)\sigma _3 \Phi (u\!+\!1,v\!+\!1)\otimes
\sigma _3\Psi (u\!-\!1,v)\,.
\end{eqnarray}
Here
\beq
\label{int}
\Phi (u,v)=\left (
\begin{array}{c} \tau (u,v,w)\\ \\
\hat \tau (u,v,w) \end{array} \right ),
\;\;\;\;\;\;\;\;
\Psi (u,v)=\left (
\begin{array}{c} \tau (u,v,w+1)\\ \\
-\hat \tau (u,v,w+1) \end{array} \right )
\eeq
($w$ is supposed to be constant), $z=\zeta / \mu$ and
\beq
\label{int1}
\begin{array}{l}
\alpha (z)=\frac{1}{2}(1+\kappa ^{-1}z^{-1})
(\kappa +1)(z-1)\,, \\ \\
\beta (z)=\frac{1}{2}(1+\kappa ^{-1}z^{-1})
(\kappa -1)(z+1)\,.
\end{array}
\eeq
Note that $\Psi$ implicitly depends on $z$ through the
shift of $w$.
In this form the identity resembles relations of the
"vertex-face correspondence" type.
It contains two linear problems whose compatibility condition
is the discrete SG equation with $\Psi$ playing the role of
the "wave function".

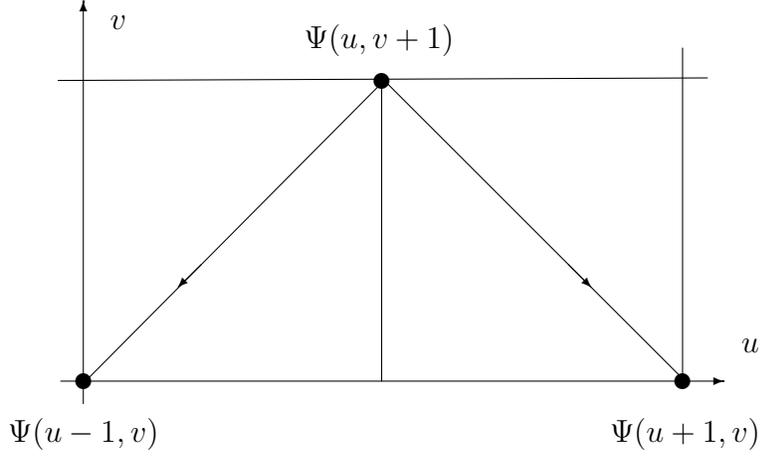
\begin{center}
\special{em:linewidth 0.4pt}
\unitlength 1mm
\linethickness{0.4pt}
\begin{picture}(98.67,70.33)
\put(95.00,20.00){\vector(1,0){0.2}}
\emline{7.00}{20.00}{1}{95.00}{20.00}{2}
\emline{6.67}{60.00}{3}{93.00}{60.33}{4}
\put(10.00,70.33){\vector(0,1){0.2}}
\emline{10.00}{17.00}{5}{10.00}{70.33}{6}
\emline{50.00}{60.00}{7}{10.00}{19.67}{8}
\emline{50.00}{60.00}{9}{90.00}{20.00}{10}
\emline{89.67}{20.00}{11}{89.67}{64.33}{12}
\emline{49.67}{60.33}{13}{49.67}{20.00}{14}
\put(49.67,60.00){\circle*{2.00}}
\put(10.00,20.00){\circle*{2.00}}
\put(89.67,20.00){\circle*{2.00}}
\put(22.75,32.60){\vector(-1,-1){0.2}}
\emline{25.65}{35.50}{15}{22.75}{32.60}{16}
\put(77.40,32.60){\vector(1,-1){0.2}}
\emline{74.51}{35.50}{17}{77.40}{32.60}{18}
\put(14.67,67.67){\makebox(0,0)[cc]{$v$}}
\put(49.33,65.33){\makebox(0,0)[cc]{$\Psi (u,v+1)$}}
\put(10.00,13.00){\makebox(0,0)[cc]{$\Psi (u-1,v)$}}
\put(90.00,13.00){\makebox(0,0)[cc]{$\Psi (u+1,v)$}}
\put(98.67,24.67){\makebox(0,0)[cc]{$u$}}
\end{picture}
\end{center}

To see this,
let us take the scalar product of both sides of eq.\,(\ref{rhir1})
with the vector $\sigma _1 \Phi (u+1, v+1)$
in the first space, orthogonal to
$\sigma _3 \Phi (u+1, v+1)$. We get:
\beq
\label{lin1}
\frac{
\Bigl ( \sigma _1 \Phi (u+1,v+1),
\,\mbox{{\sf R}}_{\kappa}(z) \Phi (u,v)\Bigr )_1}
{\Bigl ( \sigma _1 \Phi (u+1,v+1),
\, \Phi (u-1,v+1)\Bigr )}
\Psi (u,v+1)=
\alpha (z)\Psi (u+1,v)\,.
\eeq
The subscript 1 in the numerator means that the scalar product
is taken in the first space $V_1$, so the result is a
2$\times$2 matrix in the second space.
In a similar way, taking the scalar product
with the vector
$-i\sigma _2 \Phi (u-1, v+1)$, we get:
\beq
\label{lin2}
\frac{
\Bigl ( \sigma _2 \Phi (u-1,v+1),
\,\mbox{{\sf R}}_{\kappa}(z) \Phi (u,v)\Bigr )_1}
{\Bigl ( \sigma _1 \Phi (u+1,v+1),
\, \Phi (u-1,v+1)\Bigr )}
\Psi (u,v+1)=
i\beta (z)\sigma _3\Psi (u-1,v)\,.
\eeq
Note that
$$
\Bigl ( \sigma _1 \Phi (u+1,v),
\, \Phi (u-1,v)\Bigr )=2\tau (u,v)\hat \tau (u,v)
$$
due to (\ref{05}). Eqs.\,(\ref{lin1}), (\ref{lin2}) are two
linear problems for the transport of the wave function $\Psi$
in the two directions displayed in the figure. The left hand sides
of these equations provide us with the $L$-$M$-pair depending
on the spectral parameter $z$
that comes from the quantum $R$-matrix.
At the same time this spectral parameter
coincides (up to a constant scale factor)
with Miwa's variable $\zeta$.
The compatibility of the two
linear problems is expressed as a discrete
zero curvature condition equivalent to the discrete SG
equation. Given a solution to (\ref{rhir}), at
any slice $w= \mbox{const}$ we have a solution to the discrete
SG equation in bilinear form.
Therefore, we have got another proof of the results of \cite{Z},
where local $L$ and $M$ matrices for discrete
time shifts were expressed in terms of the quantum
$R$-matrix\footnote{Comparing with the final formulas of
\cite{Z}, there is a minor difference since
here we actually deal with $L^{-1}$ rather than $L$.}.

\section{Discussion}

We have shown that the $\tau$-function of the discrete SG
hierarchy apart from the standard bilinear equation (\ref{3term})
satisfies additional bilinear relations (\ref{06a})--(\ref{06c}),
the first (respectively, the last) of which provides bilinearization
for the discrete KdV equation (respectively, the 1D Toda chain in
discrete time). Besides, we have proved some additional bilinear
relations for the "extended" SG $\t$-function depending on 3 discrete
variables. They are combined into the matrix relation (\ref{rhir1})
which contains the quantum $R$-matrix of $XXZ$-type and resembles
the vertex-face correspondence. In less dry terms, this fact would
probably mean that the structure of integrable bilinear partial
difference equations is to high extent determined by solutions
of the quantum Yang-Baxter (YB) equation (\ref{YB}).

Recall that both classical and quantum YB equations already
appeared in connection with integrable mappings.
First, the (classical) $r$-matrices are known to play a significant role
in the algebraic structure of discrete
time integrable mappings \cite{Suris},\,\cite{Ragnisco}.
The quantum YB equation as arising in purely classical problems
was discussed in the works \cite{Sklyanin85},\,\cite{WX},
though the class
of relevant solutions seemed to be very far from
$R$-matrices of known quantum integrable models.
In the present paper we have observed
that a well known particular solution to the
quantum YB equation is really relevant to the discrete SG
hierarchy although in a different sense.
However, in our construction the role of the quantum YB
equation itself is not yet clear enough. Hypothetically,
it is responsible for the fact that the linear problems
(\ref{lin1}) and (\ref{lin2}) are not too overdetermined.
(In other words, the equations obtained as their consistency
conditions do not contradict to each other.)

At last,
we should stress that the "quantum parameter" $\kappa$
of the trigonometric $R$-matrix in our context
seems to have nothing to do with any kind of quantization.
It is related to the mass parameter and the
lattice spacing of the classical model.

\section*{Acknowledgements}

I thank S.Kharchev for
stimulating discussions and critical remarks.
This work was supported in part by RFBR-INTAS grant 95-1353
and by grant 96-15-96455 for support
of scientific schools.

\section*{Appendix}
\def\theequation{A\arabic{equation}}
\setcounter{equation}{0}

Here we show how to
derive the identities (\ref{06a})--(\ref{06c}) and
(\ref{e7}), (\ref{e8}).

Consider eqs.\,(\ref{03a}), (\ref{04a}) with the shift
$u \to u-1$ in the latter:
$$
\begin{array}{l}
\nu \tau (u+1,v)\hat \tau (u,v+1)
-\mu \hat \tau (u+1,v) \tau (u,v+1)
=(\nu -\mu ) \tau (u+1,v+1) \hat \tau (u,v)\,,\\ \\
\nu \tau (u-1,v)\hat \tau (u,v+1)
+\mu \hat \tau (u-1,v) \tau (u,v+1)
=(\nu +\mu ) \tau (u-1,v+1) \hat \tau (u,v)\,.
\end{array}
$$
Rewriting them as a system of linear equations,
$$
\left (
\begin{array}{lll}
\nu \tau (u\!+\!1,v) && \!-\mu \hat \tau (u\!+\!1,v) \\ && \\
\nu \tau (u\!-\!1,v) && \!\mu \hat \tau (u\!-\!1,v)
\end{array}
\right )
\left (
\begin{array}{l}
\!\hat \tau (u,v\!+\!1) \\ \\ \! \tau (u,v\!+\!1)
\end{array}
\right )
\!=\!\hat \tau (u,v)
\left (
\begin{array}{l}
\!(\nu \!-\!\mu ) \tau (u\!+\!1,v\!+\!1)
\\ \\ \!(\nu \!+\!\mu ) \tau (u\!-\!1,v\!+\!1)
\end{array}
\right )
$$
and using the Kramer rule, we get:

\beq
\label{A1}
\hat \tau (u,v+1)D=\hat \tau (u,v)
\left |
\begin{array}{lll}
(\nu \!-\!\mu )\tau (u\!+\!1,v\!+\!1)
&& \!-\mu \hat \tau (u\!+\!1,v) \\ && \\
(\nu \!+\!\mu )\tau (u\!-\!1,v\!+\!1)
&& \!\mu \hat \tau (u\!-\!1,v)
\end{array}
\right |,
\eeq

\beq
\label{A2}
\tau (u,v+1)D=\hat \tau (u,v)
\left |
\begin{array}{lll}
\nu \tau (u\!+\!1,v) &&
(\nu \!-\!\mu )\tau (u\!+\!1,v\!+\!1) \\ && \\
\nu \tau (u\!-\!1,v) &&
(\nu \!+\!\mu )\tau (u\!-\!1,v\!+\!1)
\end{array}
\right |,
\eeq

\noindent
where
$$
D=\nu \mu \Bigl ( \t (u\!+\!1,v)\hat \t (u\!-\!1,v)
+\t (u\!-\!1,v)\hat \t (u\!+\!1,v)\Bigr )\,.
$$
Taking into account eq.\,(\ref{05}), we see that
$D=2\nu \mu \t (u,v)\hat \t (u,v)$
and, therefore,
(\ref{A1}) and (\ref{A2}) coincide with
(\ref{06b}) and (\ref{06a}) respectively.
Eq.\,(\ref{06c}) is obtained in a similar way from
eq.\,(\ref{06b}) and its counterpart with exchanged
flows $u$ and $v$.

To prove (\ref{e7}), consider eq.\,(\ref{06b}) and its
counterpart with $v, \nu$ substituted by $w, \zeta$:
\begin{eqnarray}
&& (\zeta +\mu )\tau (u-1,v,w+1)\hat \tau (u+1,v,w)
+(\zeta -\mu )\tau (u+1,v,w+1)\hat \tau (u-1,v,w)
\nonumber \\
&=&2\zeta \tau (u,v,w)\hat \tau (u,v,w+1)\,.
\label{A3}
\end{eqnarray}
These
two can be rewritten as the following system of linear equations:
\begin{eqnarray}
&&\left (
\begin{array}{lll}
(\nu \!+\!\mu ) \tau (u\!-\!1,v\!+\!1,w)
&& (\nu \!-\!\mu ) \tau (u\!+\!1,v\!+\!1,w) \\ && \\
(\zeta \!+\!\mu ) \tau (u\!-\!1,v,w\!+\!1)
&& (\zeta \!-\!\mu )\tau (u\!+\!1,v,w\!+\!1)
\end{array}
\right )
\left (
\begin{array}{l}
\hat \tau (u\!+\!1,v,w) \\ \\ \hat \tau (u\!-\!1,v,w)
\end{array}
\right )
\nonumber \\
&=&2\tau (u,v,w)
\left (
\begin{array}{l}
\nu \hat \tau (u,v\!+\!1,w)
\\ \\ \zeta \hat \tau (u,v,w\!+\!1)
\end{array}
\right ).
\end{eqnarray}
Solving this system for $\hat \tau (u+1,v,w)$ and taking into
account eq.\,(\ref{e6}), we get (\ref{e7}).

To prove (\ref{e8}), consider eq.\,(\ref{06a}) with
$\t \to \hat \t$ and eq.\,(\ref{A3}) combined into
the following system:
\begin{eqnarray}
&&\left (
\begin{array}{lll}
(\nu \!+\!\mu ) \hat \tau (u\!-\!1,v\!+\!1,w)
&& (\mu \!-\!\nu ) \hat \tau (u\!+\!1,v\!+\!1,w) \\ && \\
(\zeta \!+\!\mu ) \tau (u\!-\!1,v,w\!+\!1)
&& (\zeta \!-\!\mu )\tau (u\!+\!1,v,w\!+\!1)
\end{array}
\right )
\left (
\begin{array}{l}
\hat \tau (u\!+\!1,v,w) \\ \\ \hat \tau (u\!-\!1,v,w)
\end{array}
\right )
\nonumber \\
&=&2
\left (
\begin{array}{l}
\mu \hat \tau (u,v,w)\hat \tau (u,v\!+\!1,w)
\\ \\ \zeta \tau (u,v,w)\hat \tau (u,v,w\!+\!1)
\end{array}
\right ).
\end{eqnarray}
Proceeding similarly to the previous case and using
equations from Sect.\,4, we come to (\ref{e8}).

\end{document}